# Surface-Functionalization of Oleate-Capped Nano-Emitters for Stable Dispersion in 3D-Printable Polymers

*Akhilesh Kumar Pathak[1], Sachin Prashant Kulkarni[2], Rachel R. Chan[3], Chad A. Mirkin[3], Koray Aydin[2]\*, Sridhar Krishnaswamy[1]\**

*[1]Centre for Smart Structures and Materials, Department of Mechanical Engineering, Northwestern University, Evanston, IL 60208, USA*

*[2]Department of Electrical and Computer Engineering, Northwestern University, Evanston, Illinois 60208, USA*

*[3]Department of Chemistry, Northwestern University, Evanston, Illinois 60208, USA*

*\*Corresponding: sridhar.krishnaswamy@northwestern.edu, aydin@northwestern.edu*

## Abstract

Two-photon polymerization (2PP) 3D printing is a well-known technique for fabricating passive micro/nanoscale structures, such as microlenses and inversely designed polarization splitters. The integration of light emitting nanoparticle (NP) dopants, such as quantum dots (QDs) and rare-earth doped nanoparticles (RENPs), into a polymer resist would enable 3D printing of active polymer micro-photonic devices, including sensors, lasers, and solid-state displays. Many NPs are stabilized with oleic acid ligands to prevent degradation, but oleate-capped NPs (oc-NPs) tend to agglomerate in nonpolar media despite the hydrophobicity of the ligand. This results in an uneven distribution of NPs in polymers and increased optical extinction properties. In this work, we propose a general approach for dispersing various oc-NPs in commercial 3D printable polymers. We achieve controlled growth of small carbon chains around the oc-NPs by functionalizing the NPs with methyl-methacrylate monomers. The proposed approach is validated on RENPs (~65 nm) and CdSe/ZnS quantum dots (~12 nm) using different commercial polymer resists (IP-Dip and IP-Visio). Dispersions of functionalized NPs (f-NPs) have improved NP density by an order of magnitude and are shown to be stable for several weeks with minimal impact on printing quality. Our approach is generalizable to a variety of oc-NPs and ultimately leads to higher quality polymer-based optical and electronic devices.

# 1. Introduction

Nanotechnology has revolutionized the development of active photonic devices[1], light-emitting diodes[2], optical amplifiers[3], bioimaging[4], and solid-state displays[5]. Functional nanoparticles, in particular, have proven extremely useful for realizing classical and quantum devices due to their tunable morphologies and electronic properties[6]. Among different types of NPs, rare-earth nanoparticles (RENPs) are known to display highly desirable optical properties, such as a large Stokes shift, resistance to photobleaching, and sharp emission bands[7]. Synthesizing these NPs through thermal decomposition requires oleates or acetates in boiling organic solvents, resulting in NP surfaces covered with hydrophobic ligands (e.g., oleic acid) [8,9] . The oleic acid (OA) ligand contains a long alkyl chain along with a terminal carboxyl group and plays a crucial role in the controlled growth of NPs.

The integration of NPs into dielectric media enables scalable fabrication of micro-photonic devices that surpass conventional semiconductor processes. For example, NPs with electrical, magnetic or optical properties mixed into a 3D-printable matrix can be harnessed for displays, sensors, and actuators[10,11]. To achieve a high-quality nanocomposite, it is important to maximize the loading (density) of NPs in the 3D-printable polymer matrix while minimizing changes to the photoresist's intrinsic properties. However, maximizing the loading of NPs in a polymer matrix is challenging and often leads to immediate agglomeration despite the hydrophobicity of the oleate capping. In a nonpolar matrix, agglomeration can still occur via depletion forces induced by the oligomeric components in the photoresin, leading to attractive forces between larger dopant NPs (Figure 1a)[12,13]. Large clusters formed by agglomeration result in incident light scattering and modified absorption profiles due to coupling of electronic bands between NPs [14]. Larger clusters also quickly settle to the bottom of the resin matrix, making it even more difficult to extract an evenly concentrated volume.

Several methods have been considered to improve dispersity of NPs in polymer nanocomposites. Ingrosso *et al.* developed a nanocomposite composed of an epoxy photoresist and oleate capped CdSe/ZnS quantum dots (QDs) using various liquid organic solvents to reduce the resin viscosity[15]. However, the dispersibility was limited by the polarity mismatch between the oleate ligands and the various solvents, and long-term stability was not reported. Zhao et al. synthesized a nanocomposite using RENPs dispersed in a hydrolyzed polyhedral oligomeric silsesquioxane-graft poly-methyl methacrylate (H-POSS-PMMA) matrix [16]. The resulting nanocomposite exhibited nearly single nanoparticle dispersion at a high solid loading of 10 vol% along with excellent thermal stability and optical properties, but the added processing time and chemicals significantly increased the method's feasibility and fabrication footprint. Yang *et al.* documented that full polymerization of RENPs in a methyl methacrylate (MMA) solvent generated nanocomposites with uniform and stable NP dispersions due to the unsaturated OA bonds that could copolymerize with MMA monomers (Figure 1a)[17,18]. This procedure was relatively simple and cost-effective, but the total polymerization of the mixture prevented the extraction of the NPs for use in any other polymer matrix, limiting fabrication of photonic devices to spin-coating-based methods centered around PMMA. Soon after, Momper *et al.* utilized a ligand exchange approach to remove the oleate capping from the CdSe/ZnS QDs and attach thiol-terminated poly-methyl methacrylate (PMMA-SH) chains with greater compatibility to commercial 3D-printable resins (Figure 1a)[19]. The ligand exchange process could potentially damage the NP surface as the protective ligand is removed and a new ligand is added, making it a risky approach for highly sensitive quantum emitters[20]. The process is also limited to NPs composed of materials with a high affinity for thiol bonds (plasmonic NPs, sulfide QDs, etc.), effectively excluding $NaYF_4$-based RENPs.

A simple procedure that does not remove the native ligand in conjunction with the ability to transfer processed NPs to other resins would be highly desirable for 3D-printing of active photonic devices. We hypothesize that by intentionally terminating the polymerization of an NP/MMA mixture before completion, NPs can be functionalized with PMMA chains over the oleic acid ligand, allowing extraction and re-suspension of NPs into various 3D-printable polymers with more uniform and stable dispersion compared to unmodified NPs (Figure 1b). While unmodified NPs require mechanical dispersion and begin to settle after a few days, our functionalized NPs exhibit great dispersibility and long-term stability for several weeks. Our approach has been validated for various sizes and types of photoluminescent NPs, with oleic acid ligands (QDs and RENPs). The improved dispersibility, long-term stability, and generalizability of this process makes it possible to synthesize uniform polymer nanocomposites for 3D-printing of optical devices.

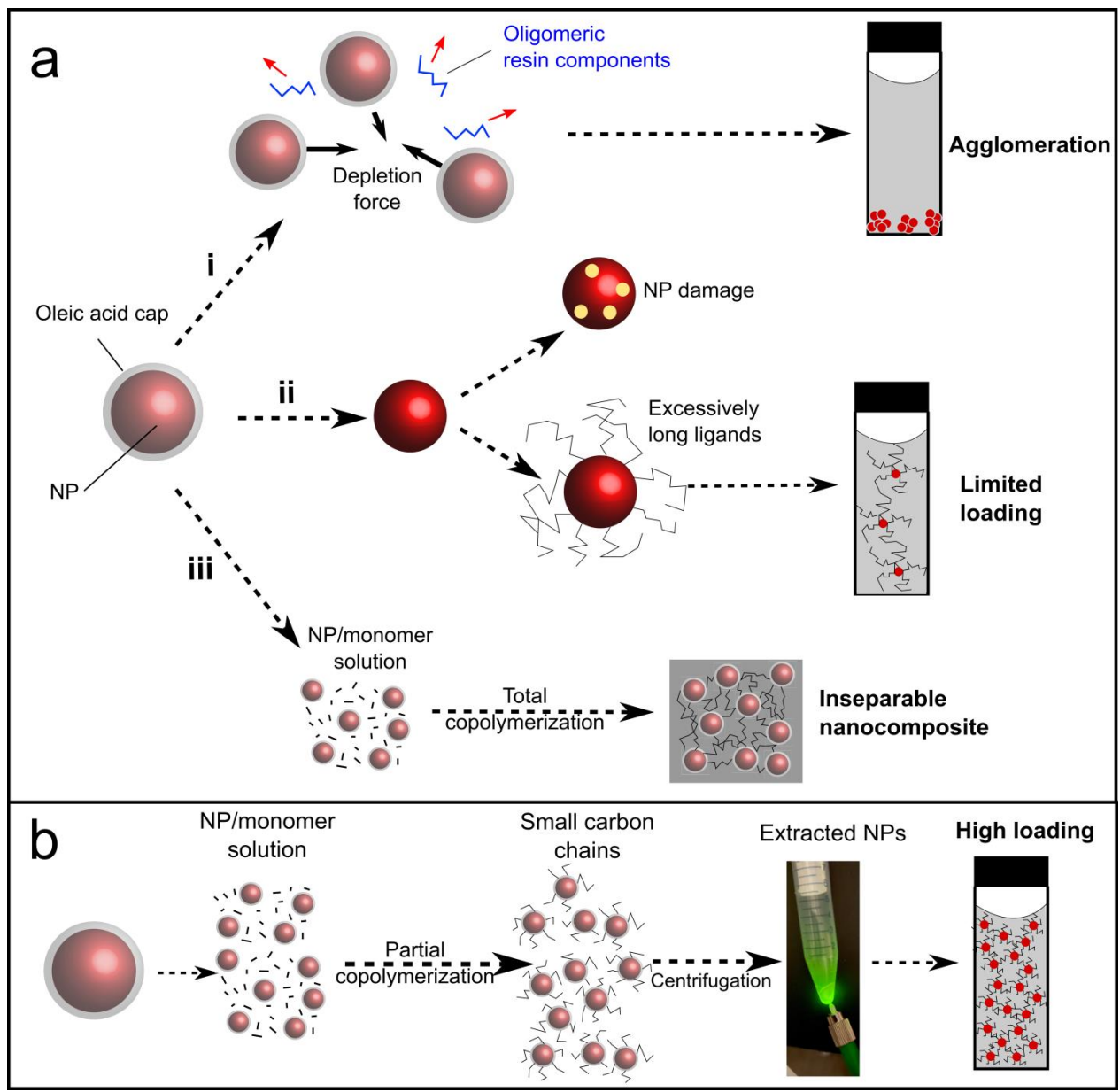

**Figure 1** *a) Consequences of various dispersion methods explored in existing literature, such as i) direct dispersion of oc-NPs, ii) ligand exchange, and iii) nanocomposite formation. b) a schematic of our new procedure that enables NP functionalization and dispersion in other resins.*

## 2. Results and discussion

### 2.1. Functionalization and Process Optimization

The dispersibility of the oc-NPs depends on the compatibility of the ligand covering the NPs with a polymer matrix. Both oleate-capped RENPs (Er,Yb:NaYF$_4$) and quantum dots (CdSe/ZnS, 620nm emission) are considered as candidates for surface modification. Core/shell RENPs were synthesized according to standard procedures reported in the literature (Supplementary Information S1). The synthesized RENPs are hexagonal in shape with sizes ranging from 50 to 65 nm (Supplementary Information S2). Commercially available QDs are spherical with particle sizes ranging from 8 to 12 nm (Supplementary Information S2).

For attaching MMA to the OA ligand, our process is derived from the procedure described in Yang et al. However, the reaction time and temperature are adjusted to control the extent of MMA polymerization (See Methods section). The mixture fully polymerizes and solidifies after 2 hours, so the reaction is studied within this timeframe. To observe the degree of agglomeration during the reaction, a small amount of the mixture is extracted each hour and drop-cast onto a glass slide for fluorescence imaging. The fluorescence images from samples extracted at 0 hours and 2 hours of processing (Figures 2a and 2b) are shown with corresponding emission spectra measured from various locations within the area (Figures 2c and 2d). The variability in emission decreases in the sample processed for 2 hours, whereas the unprocessed sample generates micron-scale clusters with extremely strong emission from these sites and weaker emission from the background. The energy-dispersive x-ray (EDX) spectra of oc-RENPs and functionalized-RENPs (f-RENPs) are shown with their corresponding TEM images (Figures 2e and 2f). The spectra reveal expected peaks corresponding to the RENP constituent elements, including sodium (Na), yttrium (Y), fluorine (F), ytterbium (Yb), and erbium (Er). However, increased carbon and oxygen peaks for the functionalized RENP sample validate the growth of carbon chains on f-RENPs. Although a slight degree of nanoparticle agglomeration can be observed in both TEM images, the f-RENPs form much smaller clusters when removed from the solution. The aggregation of RENPs in PMMA at higher loadings has been previously reported, but in our case, they maintain their dispersity at higher loadings [21].

The progress of the polymerization reaction is quantified using Fourier Transform Infrared (FTIR) spectroscopy. To monitor the effect of the degree of polymerization, 1 ml of the polymerized mixture is extracted every 30 min. The FTIR spectra of pure MMA and RENPs functionalized with PMMA for 120 min are shown below (Figure 2g). The peak marked at $1610\,cm^{-1}$, corresponding to the stretching vibration of the C=C in the spectra of pure MMA, implies the presence of the monomer[22]. During polymerization, the C=C bond is broken to form C-C bonds with neighboring monomers. Consequently, the peak corresponding to the C=C stretching vibration at $1610\ cm^{-1}$ vanishes as copolymerization is allowed to progress, indicating the growth of a carbon chain. The complete FTIR analysis with all measurements from the intermediate time intervals is also shown (Supplementary Information S3).

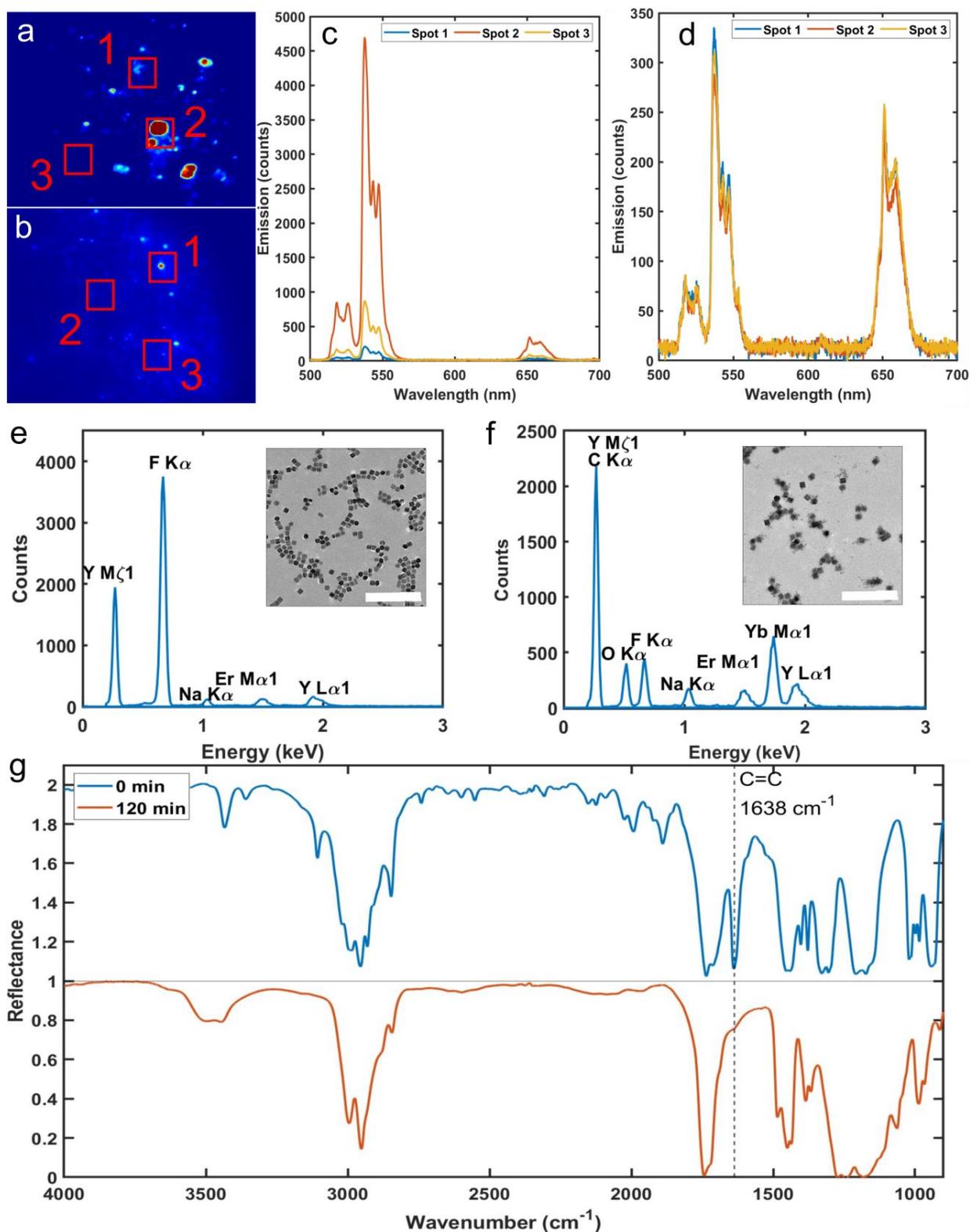

***Figure 2*** *Characterizing polymerization progress over time. a,c,e) Upconversion fluorescence imaging after drop-casting, emission spectra for the marked locations, and the EDX spectrum with a TEM image of the MMA/RENP after 0 minutes of processing. b,d,f) Upconversion fluorescence imaging after drop-casting, emission spectra for the marked locations, and the EDX spectrum with a TEM image of the MMA/RENP after 120 minutes of processing. The scale bar on the TEM insets is 200 nm. g) FTIR analysis of the MMA/RENP*

*solution before and after 120 minutes of processing showing the vanishing peak at 1638 cm$^{-1}$, indicating polymerization.*

## 2.2. Dispersibility/Stability of Nanoparticles in 3D-Printable Resins

Once the polymerization reaction is optimized, the f-NP dispersibility needs to be validated in a liquid photoresin. Two types of functionalized oleate-capped photoluminescent nanoparticles (RENPs and commercial CdSe/ZnS QDs) were modified and redispersed in resins to assess dispersibility, stability, and functionality. The distribution of nanoparticles in a resin can be qualitatively gauged by the transparency of the mixture, which is essential for propagation of the pump laser in two-photon polymerization (2PP) printing. Oc-RENPs, f-RENPs, oc-QDs, and f-QDs and commercial CdSe/ZnS quantum dots were mechanically dispersed into proprietary photoresins (IP-Dip and IP-Visio, Nanoscribe GmbH). Changes in resin transparency are also documented before and after the addition of oc-NPs and f-NPs (Figure 3a). While the f-NPs maintain the transparency of the polymer, the oc-NPs scatter the ambient light to generate a cloudy translucent resin, which is most clearly observed in the oc-RENP mixture. In the case of the oc-QD mixture, the agglomeration of particles is severe enough that individual clusters can be visibly observed in the resin even after sonication for more than 12 hours. Despite having the same concentration, there is a stronger color imparted to the oc-QD mixture compared to the f-QDs mixture, likely due to the optical scattering properties of clusters comparable to visible light wavelengths.

Fluorescence imaging of the oc-QDs in photo resin shows non-uniform dispersion with bright clusters and dark regions in between (Figure 3b). However, with f-QDs, fluorescence throughout the resin is more uniform with no discrete clusters and less emission contrast (Figure 3c). The dispersion is also gauged quantitatively by curing a drop of each mixture on a glass microscope slide using a UV lamp and measuring the emission spectrum from various locations (Figures 3d-3g). The oc-NPs show a large variation in the emission intensity due to agglomeration of emitters, whereas the f-NP emission is uniform at different locations.

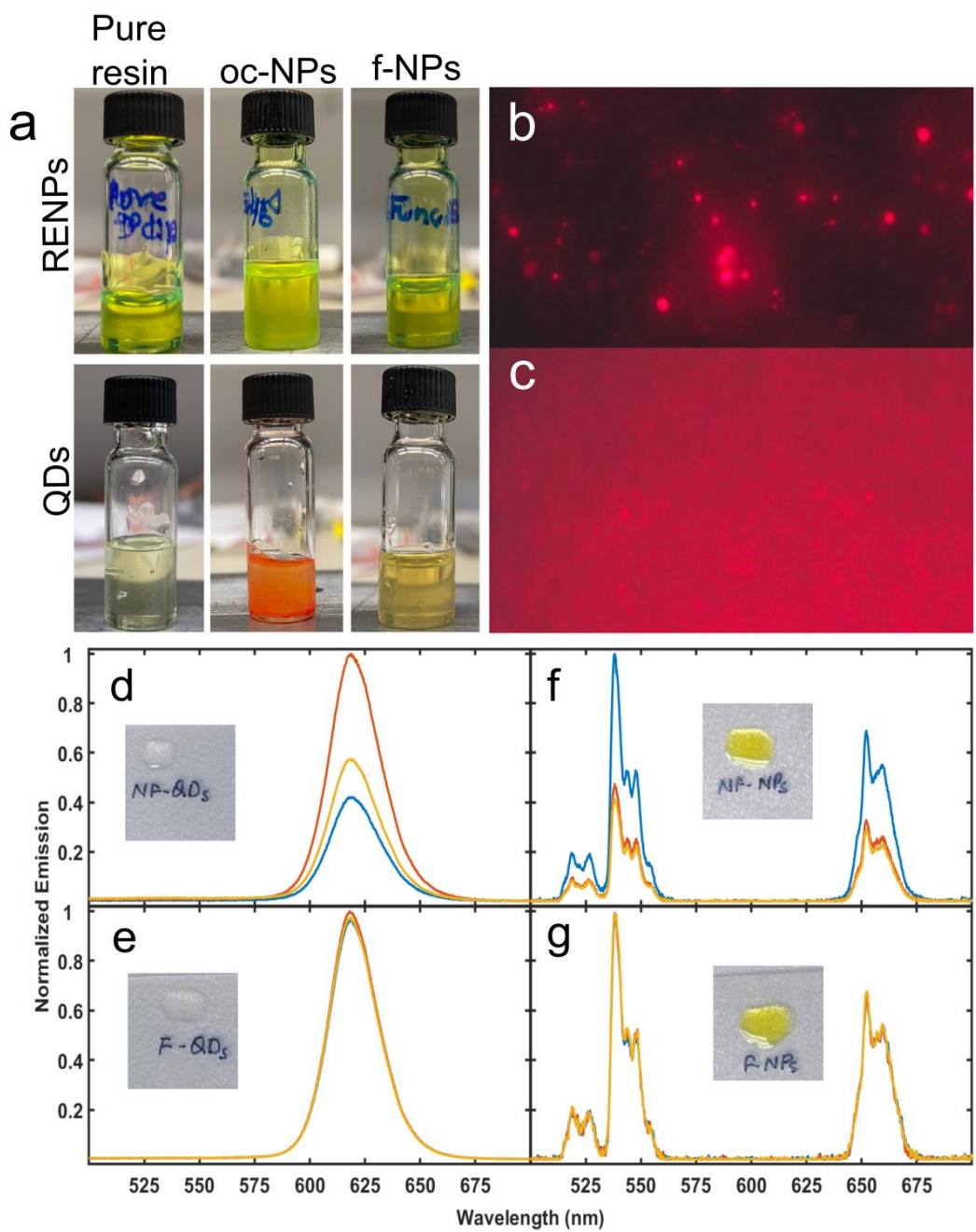

***Figure 3*** *Dispersion of NPs in liquid IP resin. a) Vial images showing the color and transparency of the photoresist in the pure state (left), with oc-NPs (middle), and f-NPs (right) for RENPs (top row) and QDs (bottom row). b,c) Fluorescence images of oc-QDs (b) and f-QDs (c) after being mixed into IP-Visio. d-g) Emission spectra for the oc-QD (d), f-QD (e), oc-RENP (f), and f-RENP (g) composites. The insets show the cured composites after being dropped onto a glass slide. The different colors on the plots indicate different locations from where emission was measured.*

To validate the dispersion of nanoparticles in low-volume structures after printing, we use two-photon 3D printing to fabricate square slabs with side lengths of 200μm and heights of 1.5μm and map the distribution of NPs by taking images while scanning across the area with the pump laser (Supplementary Information S4). The height is deliberately kept small to ensure that most nanoparticles within the slab remain in focus during imaging and that the pump is not depleted through the height of the structure. There is a clear improvement in nanoparticle loading for both the f-QDs and f-RENPs. Since fluorescence imaging of NPs is diffraction-limited, it is impossible to differentiate between individual emitters and small clusters that still fall under a single pixel. Therefore, emitter-density calculations must be generalized to the number of emitting-sites (ESs) rather than individual particles. By determining the number of ESs within printed structures, we expect that f-NPs will have a higher ES density compared to oc-NPs. The ES density for oc-QDs is $1.6E15$ m$^{-3}$ and increases to $8.05E15$ m$^{-3}$ for f-QDs. Similarly, the RENPs see a density increase from $3.05E15$ m$^{-3}$ to $1.19E16$ m$^{-3}$. It is expected that the QDs would see a more dramatic improvement in loading due to their smaller size in comparison to the RENPs. With a statistical analysis of the mean number of individual particles per ES, it is possible to convert the ES density to the nanoparticle density, but this would require a characterization method capable of resolving individual NPs within the polymer slab. However, with the data shown, it can only be claimed that the QD density is at least $8.05E15$ m$^{-3}$. The erbium ion density with f-RENPs was estimated to be $4.63E20$ m$^{-3}$ (Supplementary Information S5), which is comparable to the standard for Er-doped optical fibers used in amplifiers ($1E25$ m$^{-3}$ - $1E26$ m$^{-3}$) [23,24]. The efficiency of such devices heavily relies on the emitter loading. Lower loading of emitters inherently generates a weaker output, but excessively high loading can also suppress output emissions by means of concentration quenching and other dissipative pathways[25].

To further characterize the uniformity of each distribution, the mean width of the ESs in each sample is also extracted from NP dispersion maps. The mean number of pixels for each ES is reduced for both types of functionalized NPs, by 40% for the QDs (3.77 pixels/EC to 2.45 pixels/EC) and 58% for the RENPs (7.34 pixels/ES to 3.08 pixels/ES). Minimizing the mean size of ECs is essential to the development of polymer-based emitting displays, where a finer dispersion of NPs enables higher quality color mixing using different QD sizes as well as a more evenly lit pixel, which can improve the total image quality.

The stability of the NPs' dispersion in photo resin is also critical to their printing performance and commercial viability. In order to achieve uniform nanocomposites, it is essential that NP dispersions can be sustained for long periods of time. The stability of each NP dispersion is studied for multiple weeks for both NP/photoresist mixtures. In the case of the oc-NPs, 70 mg of oc-RENPs and 0.25 mg of oc-QDs are mechanically dispersed in IP Dip and IP Visio, respectively, without any modifications. The nanocomposites are left undisturbed for the duration of the study without any sonication or physical shaking, and the fluorescence of each vial under illumination by a pump laser is imaged at 0 days and 14 days after functionalization. Visual assessment of laser propagation quality and emission distribution at various depths of the oc-NP and f-NP mixtures is compared (Figures 4a and 4b). In a cloudy mixture, the laser is either scattered or absorbed to the point that the beam is dissipated before reaching the other end of the vial, while the clearer vials enable the laser to pass through as a relatively undisturbed beam. Both functionalized and non-functionalized particles are stable on the first day of dispersion with uniform emission distributions (Figure 4a). However, after two weeks, uneven emission distribution is observed throughout the vial for oc-RENPS and oc-QDs, with less emission at the top and more scattering at the bottom. In contrast, uniform emission is clearly visible throughout the vial for f-RENPs and f-QDs, indicating a stable distribution of both RENPs and QDs in their respective 3D printable resin. Emission spectra were also recorded for both f-NPs at the same time points (0 and 14 days). The spectra were recorded at three different points on the sample and show a minimal drop in emission after two weeks (Figure 4c). In the oc-RENP mixture, there are pronounced changes in the fluorescence intensity as the RENPs settle down over time, but in the f-RENPs, the fluorescence shape is relatively consistent over the duration of the test. Even though the confinement of fluorescence around the pump beam path is clear for the two-photon fluorescence of the RENPs, the single-photon fluorescence of the QDs allows the beam to excite a wider radius despite the beam waist being smaller. This is largely because the emission spectrum of QDs partially overlaps with the excitation spectrum, allowing for neighboring QDs to be excited by the emissions from QDs in the pump beam path. Therefore, the beam appears diffuse even in the better dispersed f-QD mixture. The agglomeration in the oc-QDs is emphasized by the brighter fluorescence on the right side of the vial where the pump beam is located. The fluorescence intensity drastically decreases over the length of the vial, showing that a significant amount of pump power is dissipated through absorptive bands and scattering. Nanocomposites composed of f-NPs in printable photo resins are promising material candidates in polymer-based active photonic devices due to their superior stability.

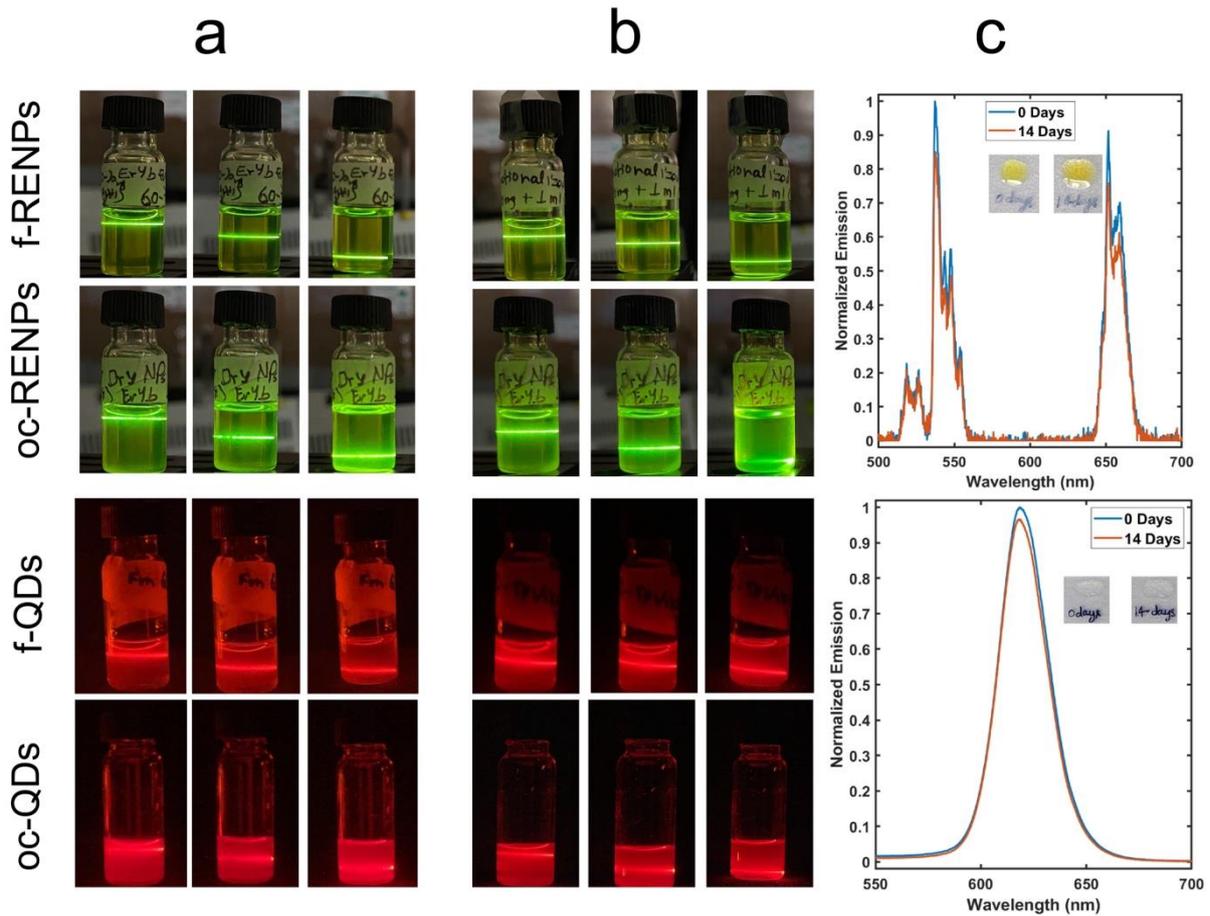

***Figure 4*** *Stability of f-NPs and f-QDs in IP-Dip and IP-Visio, respectively. (a) initial stability of f-NPs and f-QDs on day 1 using NIR and green lasers for RENPs and QDs, respectively, the emission is compared with oc-RENPs and oc-QDs vials (b) stability after two weeks showed uneven emission in oc-RENPs and oc-QDs compared to the emission from f-RENPs and f-QDs vials (c) Emission spectra over two weeks with a minimal drop in emission averaging three points.*

## 2.4. Printed Structures with Functionalized Nanoparticles

In 2PP-printing, polymerization is achieved through two-photon absorption of coherent near-infrared (NIR) photons, which can propagate through polymer resins at lower intensities[26,27]. Tightly focusing the NIR beam creates a small volume with sufficient power at the focal point for absorption-induced polymerization to occur, while leaving the surrounding volume unaffected (Figure 5a). The resolution capabilities of 2PP-printing are defined by this volume, known as the voxel. While printing, the voxel often overlaps with the substrate to ensure adhesion of printed structures. To ensure that the voxel reaches the surface of the substrate, the objective lens is immersed in the photoresist droplet, enabling the use of objectives with lower working distances and higher numerical apertures. After the desired structures are printed, the excess liquid resin is washed away using an organic solvent, like propylene glycol methyl ether acetate (PGMEA), followed by an additional wash with isopropanol (IPA).

To highlight the effect of improved dispersibility in printed structures, some example designs are printed using the f-RENP and f-QD mixtures. Momper *et al.* highlights how scattering and absorption from NP clusters leads to substantial deformations or resin degradation in printed structures [14]. However, our prints show that f-NP photo-resin composites achieve printing qualities comparable to unmodified resins (Figures 5b-5g). To similarly test the functionalization method with a structure involving finer features, the Northwestern Wildcat logo is printed over an area of ~200µm x 200µm with the smallest features, like the whiskers, having a width of approximately 500nm as measured by SEM (Supplementary Information S6). It is evident that even sub-micron features can still be printed with the same quality as the pure polymer. Both the f-RENP and the f-QD prints of the N-structure are also shown (Figures 5h and 5i). A representative QD wildcat image is shown because QDs can be pumped over a total area with a substantially lower power compared to RENPs (Figure 5j). From fluorescence alone, the desired images are well-resolved despite the small regions with lower concentrations of nanoparticles. Both types of f-NPs are mostly contained within the smaller features of the prints, indicating that the functionalization method can be utilized for a variety of micro-photonic applications. A comparison of oc-NP prints and f-NP prints highlights the large difference in image quality, as even after multiple iterations, the oc-NP print loading was far lower than the f-NP prints (Supplementary Information S7).

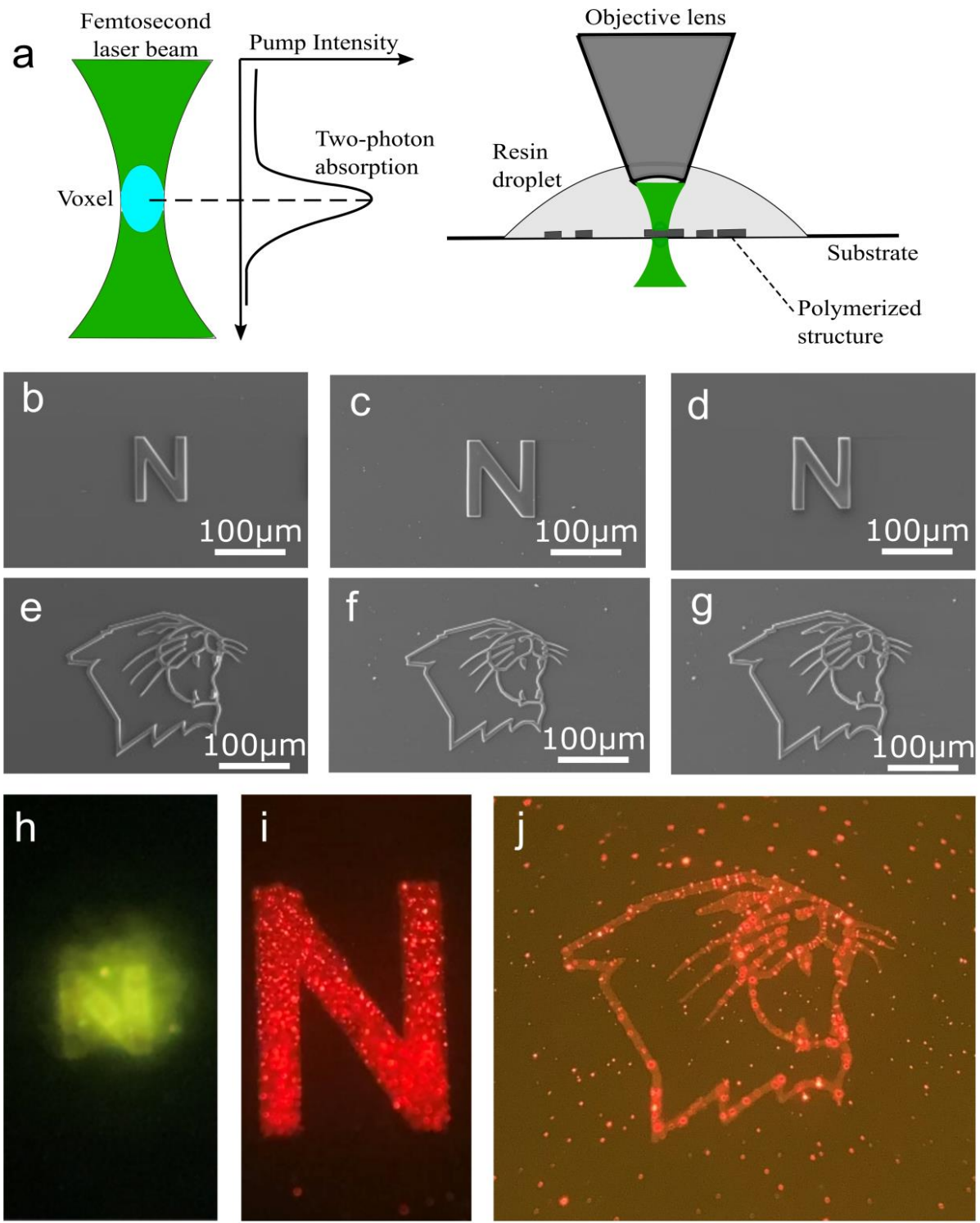

**Figure 5** *Overview and analysis of printed structures using composite resins. a) A schematic of the 2PP-printing process. b-g) SEM images of the printed N (top) and wildcat (bottom) designs using pure resin (left), f-RENPs (middle), and f-QDs (right). h,i) Fluorescence images of the N structures for f-RENPs (h) and f-QDs (i). j) A fluorescence image of the wildcat design using the f-QD composite.*

Overall, our functionalization method demonstrates a strong improvement to nanoparticle dispersibility in 3D-printable resins. However, certain challenges remain, such as the adhesion of f-NPs to the substrate in areas without printed structures. While this is relatively inconsequential for gain-based photonic devices, the adhesion of emitters to unwanted locations can be problematic for solid-state displays, where the confinement of emitters to designated regions directly affects image quality. The partial copolymerization process also warrants further investigation regarding whether the interparticle spacing could be controlled using the processing duration, as this would directly affect the NP loading. This would be especially relevant for 3D-printed lasers, amplifiers, and other photonic devices where gain is determined by NP loading.

## 3. Conclusion

In this work, we have developed a surface-functionalization method for oleic acid capped NPs that reduces agglomeration in nonpolar polymers. Our method functionalizes short PMMA chains alongside native oleic acid ligands onto the surface of NPs. This allows for separation and re-suspension of functionalized NPs into a variety of photoresins and circumvents the need for ligand stripping. Our method is validated on different fluorescent NPs with sizes ranging from 10nm to 60nm using multiple resins compatible with 2PP printing. Functionalized NP suspensions remain stable over longer time periods (at least 14 days) compared to oc-NPs. Using these nanocomposite resins, printed structures with varying feature size show that printing quality is not compromised by the introduction of f-NPs. Ultimately, by providing a route towards uniform NP dispersion in photoresins, this work makes an important step towards increasing the resolution and quality of 3D-printed nanostructures for active photonic and optical applications.

## Methods:

**Materials:** $ErCl_3.6H_2O$, $YbCl_3.6H_2O$, $YCl_3.6H_2O$, oleic acid, 1-octadecene, sodium hydroxide, ammonium fluoride, methanol, n-Butyl acetate, methyl methacrylate, 2,2'-azobis-isobutyronitrile are purchased from Sigma Aldrich. All chemicals were of analytical grade and utilized without further purification. Commercial photo-resins (IP Dip and IP Visio) are obtained from Nanoscribe GmbH. The quantum dots were acquired from NNCrystal US Corporation under the product number HECZ620-25.

**Synthesis of oc-RENPs:** The core/shell RENPs have been synthesized using thermal decomposition method from previously reported articles[28,29]. The SI offers a detailed explanation of the synthesis process.

**Carbon chain growth:** The chemical composition is adapted from the article reported by Yang *et al.*[18]. In a single-necked flask, 10 g of MMA are combined with 2 ml of n-Butyl acetate and 0.01 g of 2,2'-azobis-isobutyronitrile are utilized to grow carbon chain on NPs (as shown in Fig. 1b). Following pre-polymerization at 70 °C for 2 hours, 4 ml of n-Butyl acetate containing 0.2 mmol of oc-NPs (RENPs and QDs) is added into the flask dropwise. Subsequently, the temperature of the flask is increased from 70 °C to 95 °C for copolymerization. The temperatures for the pre-polymerization and copolymerization steps have been adjusted from the published procedure, with 70 °C for pre-polymerization and 95 °C for copolymerization observed to give the desired growth of the carbon chain. At 120 min, the f-NPs are retrieved by centrifugation of the nanocomposite at 3000 rpm for 15 minutes.

**Dispersion of f-NPs in resin:** All 3D printable resins are commercial and are used without any purification or modification. The f-NPs and directly dispersed in the IP resist. After adding the f-NPs, the mixture is sonicated for uniform dispersion. The f-RENPs and f-QDs both required 20 minutes of sonication for uniform dispersion. More details about the particle size, concentration, and dispersion are shown in the following table.

| Particles | Size | Used 3D printable resin | Concentration | Sonication |
|-----------|------|-------------------------|---------------|------------|
| RENPs | 50-65 nm | IP-Dip (1 ml) | 60 mg/ml | 20 minutes |
| CdSe/ZnS QDs | 8-12 nm | IP-Visio (1 ml) | 2.5 mg/ml | 20 minutes |

**2PP printing parameters:** 2PP-printing was performed using the NanoScribe Photonic Professional (GT) 3D-printer. The structures were printed inside a small droplet of 3D printable negative photoresin (IP-Dip/IP-Visio) using dip-in laser lithography. A 25x N.A immersion objective was used to focus the laser at voxel point, and a standard glass microscope slide was used as the substrate. After printing, the excess resin was rinsed off of the structure via immersion in propylene glycol methyl ether acetate (PGMEA) for 15 minutes, followed by immersion in isopropanol (IPA) for 5 minutes. The printing parameters, such as scan speed and laser power, are optimized for the resin type and the structure morphology.

**Optical Measurements/Imaging:** All emission spectra were collected using a Nikon Ti-U inverted microscope connected to a cooled Andor Newton 971 EMCCD camera through an Andor SR-303i monochromator. The pump sources were a PicoQuant LDH-P-C-375 laser for the quantum dots and a PicoQuant LDH-D-C-975 laser for the RENPs. For the stability images involving the pump laser, the QD vials were illuminated with a PicoQuant LDH-P-C-510 instead of the UV laser to prevent the IP-Visio from polymerizing. NP dispersion maps were constructed by scanning the elliptical beam spot of each laser across the sample area using a translation stage as a laser mount and stacking the resultant images. Both the emission spectra and the NP dispersion maps were collected using a Nikon 20x/0.4 objective lens. Fluorescence images for the QD samples were taken using a Zeiss Axio Observer Z1.m inverted microscope. The pump source was a broadband tungsten lamp with a 450nm long-pass dichroic mirror to isolate the pump wavelengths.

## Acknowledgements

A.K.P. and S.P.K. contributed equally to this work. S.K. acknowledges support from the Office of Naval Research through award numbers N00014-23-1-2529 and N00014-21-1-2233. K.A. and C.A.M. acknowledge support from the Air Force Office of Scientific Research under award number FA9550-22-1-0300. R.R.C. gratefully acknowledges support by the National Science Foundation Graduate Research Fellowship Program under grants DGE-2234667. Any opinions, findings, and conclusions or recommendations expressed in this material are those of the authors and do not necessarily reflect the views of the National Science Foundation. This work made use of the two-photon 3D printer at the Center for Smart Structures and Materials at Northwestern acquired with support from the ONR through award number N00014-15-1-2935. This work made use of the EPIC and BioCryo facilities of Northwestern University's NUANCE Center, which has received support from the SHyNE Resource (NSF ECCS-2025633), the IIN, and Northwestern's MRSEC program (NSF DMR-1720139).

**Supporting Information**

**1. Synthesis of core/shell oleate-capped rare earth nanoparticles:** We synthesize 50-65 nm core/shell RENPs ($NaYF_4$:$Er^{3+}$ $Yb^{3+}$@$NaYF_4$). The cores of RENPs have been synthesized using the thermal decomposition method[28]. All the rare-earth-metal chlorides $ReCl_3$.$6H2O$ (Re = Er = 2%, Ce =6%, Yb = 18%, Y=74%) are mixed in 6 ml of oleic acid and 15 ml of 1-octadecene in a three-neck flask. The mixture is degassed for 2 hours using argon gas, followed by heating at 100 ℃ for 10 min and 150 ℃ for 30 min until a homogenous transparent solution is obtained. After heating, the solution is allowed to cool down to room temperature in the presence of argon. A separate mixture containing 0.1 g of NaOH and 0.1482 g of $NH_4F$ in 10 ml of methanol is added in single drops for 30 min, followed by an additional three hours of degassing. The degassed mixture is heated at 110 ℃ for 30 min and later at 310 ℃ for 1 hr. The temperature is maintained throughout the process, as it plays a crucial role in the phase transformation of the $NaYF_4$ matrix. While approaching 310 ℃, the argon flow is removed at ~180 ℃. All outlets are sealed with a rubber septum, and the heating is continued. The temperature of the solution is monitored periodically using a thermocouple. When the reaction is complete, the mixture is cooled down to room temperature. 70 mL of ethanol is poured into the flask to precipitate the nanoparticles. The mixture is centrifuged at 3000 rpm for 15 min to retrieve the nanoparticles. The nanoparticles are further washed with ethanol thrice and dispersed in cyclohexane (which is known as the parent core). To synthesize the shell on top of this core, we utilize a similar approach to the one discussed above. 1 mmol of $YCl_3$•$6H_2O$ is mixed with 6 ml of oleic acid and 15 ml of 1-octadecene in a three-neck flask. The steps above are repeated until a homogenous transparent solution is achieved. After cooling the mixture to room temperature, the nanoparticle cores dispersed in cyclohexane are added to the mixture. The solution is then heated to 110 °C under $N_2$ atmosphere to evaporate cyclohexane from the solution. After removing the cyclohexane, the synthesis procedure is the same as that of the nanoparticle cores. Finally, the core/shell nanoparticles are retrieved using multiple centrifugations and dispersed in cyclohexane for future use.

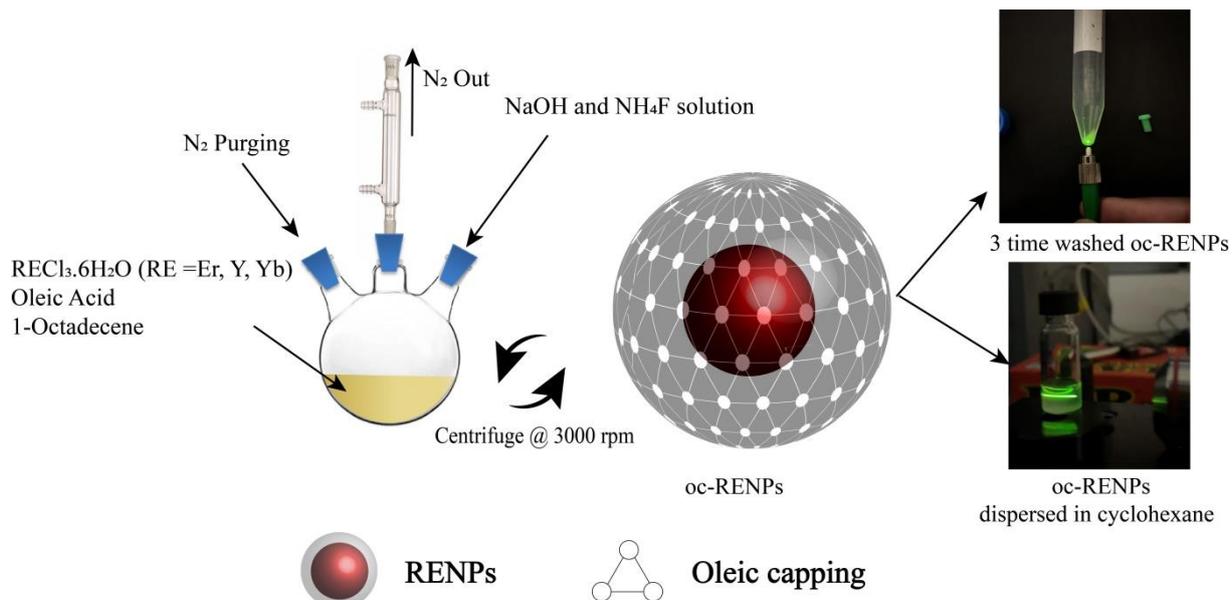

**Fig. S1.** *Schematic for the synthesis of oc-RENPs*

## 2. NPs size distribution and emission spectra:

The TEM micrographs display synthesized core/shell oc-RENPs and commercially acquired oc-QDs (Fig. S2 a and b, respectively). The oc-RENPs exhibit a hexagonal morphology with particle sizes ranging from 50-65 nm, as depicted in Fig. S2c. In contrast, the commercially obtained oc-QDs appear spherical, as demonstrated in Fig. S2d with particle size ranging between 8-12 nm. Emission spectra for oc-RENPs and oc-QDs are illustrated in Fig. S2(e and f), respectively.

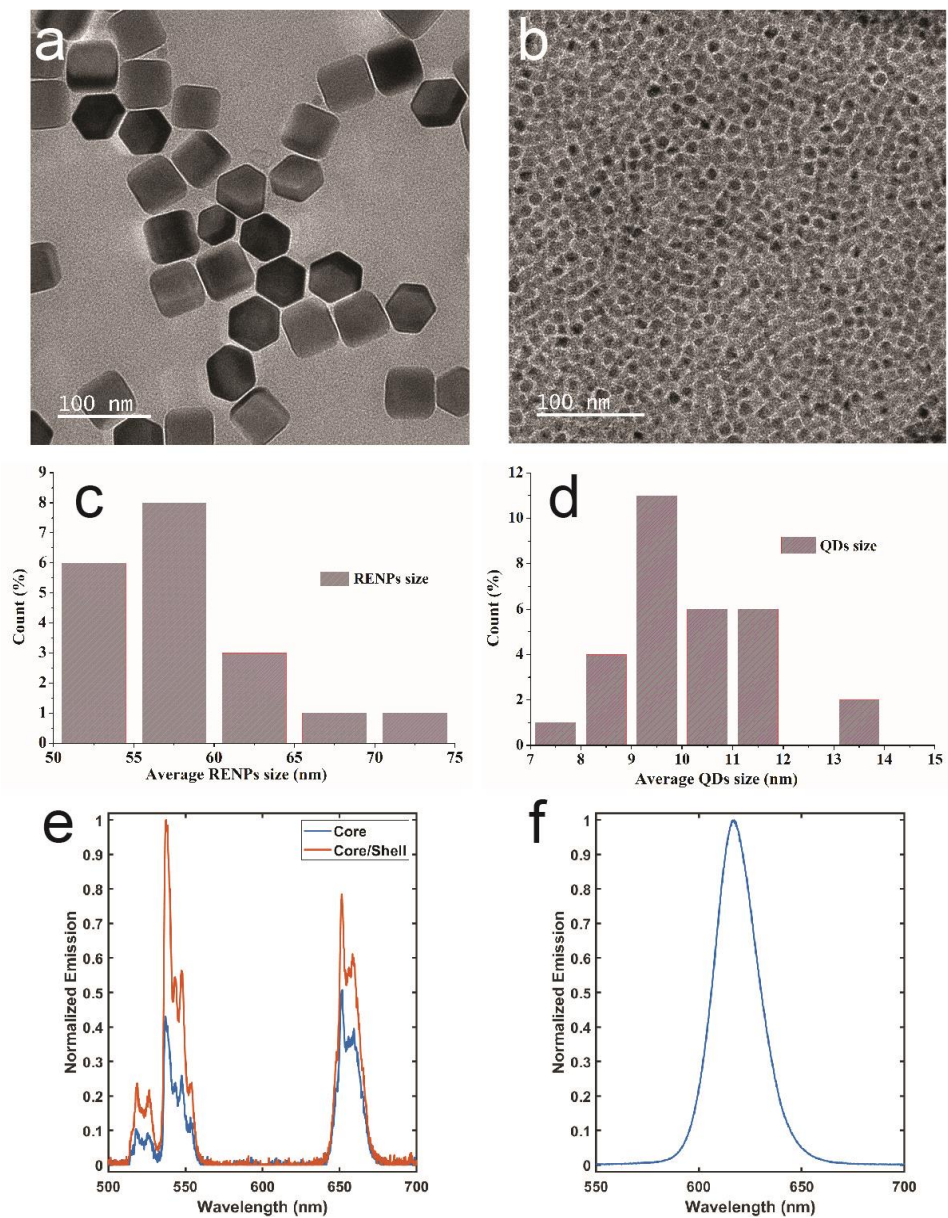

**Fig. S2.** *Size and emission characterization of RENPs (a,c,e) and QDs (b,d,f).*

## 3. FTIR Analysis:

A sharp and intense peak appears at 1730 cm$^{-1}$ in pure MMA due to the stretching vibration of the ester carbonyl group (C=O) in the MMA structural unit of the polymer. The broad band around 1643–1592 cm$^{-1}$ is related to the O–H bending and the C=C vibrations of MMA. Both spectra show absorption bands around 2854 cm$^{-1}$ and 2924 cm$^{-1}$, which are related to the C–H symmetric and asymmetric stretching modes, respectively. The FTIR spectra of RENPs functionalized with PMMA exhibit changes in several absorption peaks compared to the pure MMA sample. The peaks around 2854 cm$^{-1}$ and 2924 cm$^{-1}$ are assigned to the stretching vibration of CH$_3$ and CH$_2$. The intense peak around 1730 cm$^{-1}$, in contrast to MMA, is associated with the stretching vibration of the C=O group of PMMA. Other peaks around 1271, 1240, 1193, and 1150 cm⁻¹ correspond to the stretching vibration of C–O–C unit in PMMA.

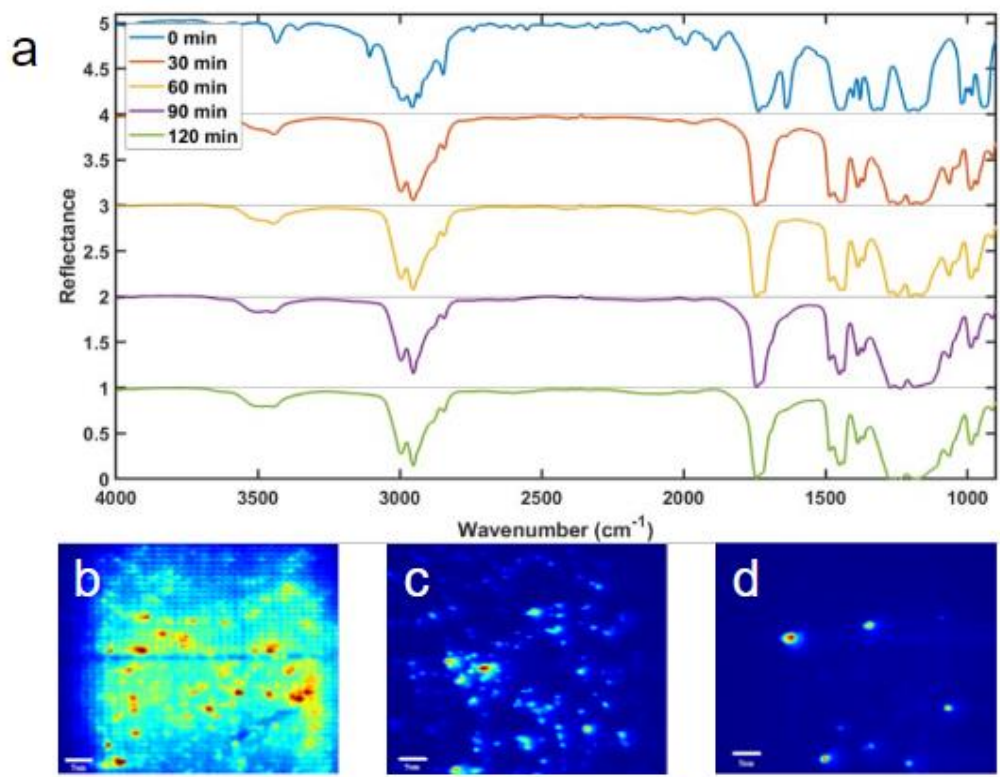

***Figure S3*** *Full FTIR analysis and fluorescence imaging for RENPs as a function of polymerization time. a) FTIR reflectance of the MMA/RENP mixture taken every 30 minutes. b,c,d) Fluorescence images of the drop-casted MMA/RENP mixture on a glass slide at 0, 60, and 120 minutes of processing, respectively.*

## 4. NP dispersion images for Emitting-Site (ES) calculations

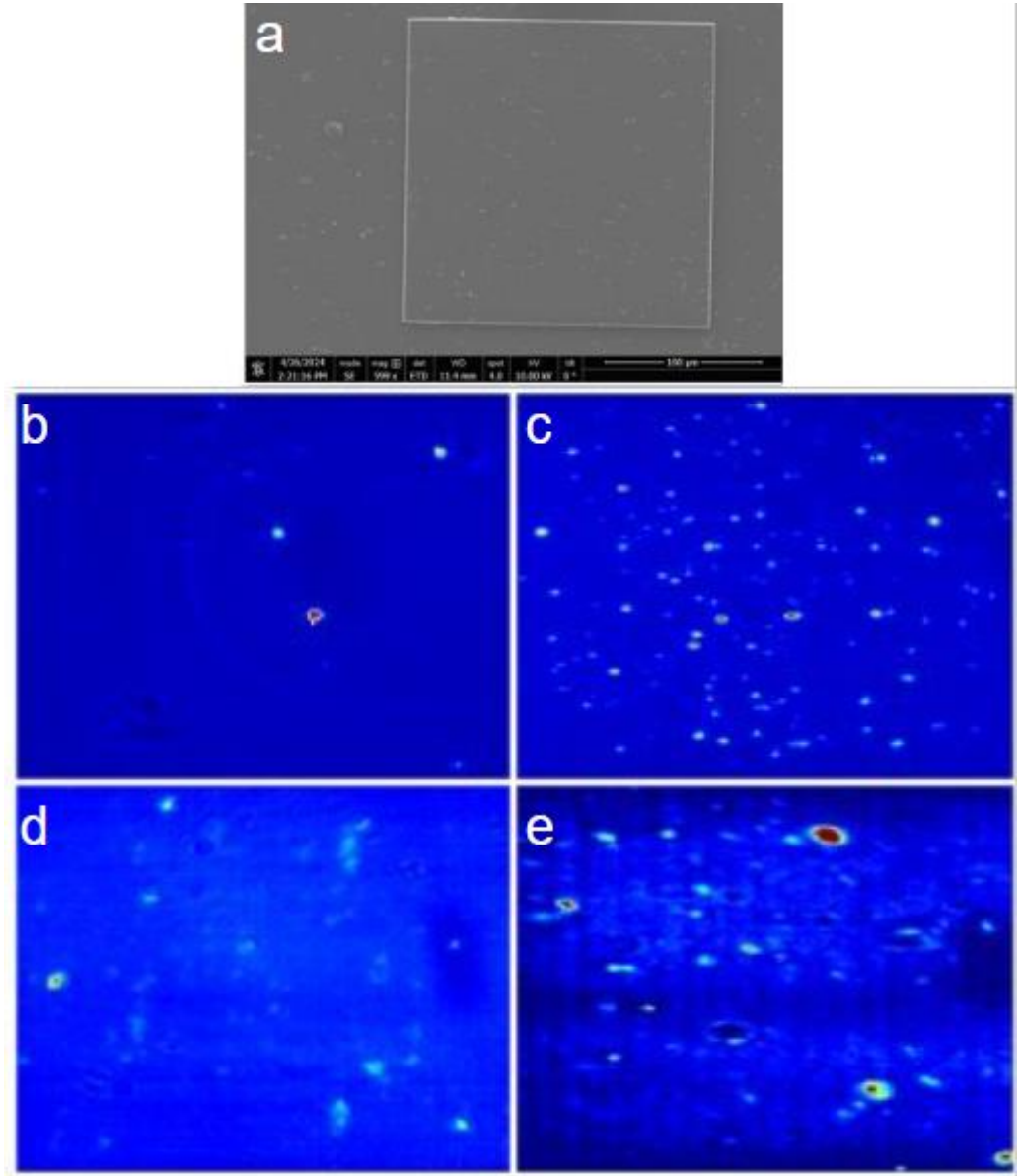

***Figure S4*** *SEM (a) and fluorescence images for the printed slabs with oc-QDs (b), f-QDs (c), oc-RENPs (d), and f-RENPs (e).*

**5. Calculations of Er$^{3+}$ ions from ES:** To perform the calculation, we assume the oc-RENPs to be spherical in shape, allowing us to define their volume as:

$$V_{oc-RENPs} \; = \; \frac{4}{3}\pi r^3$$

where the radius of oc-RENPs is 28 nm (d = 56 nm), as measured from the TEM data in Fig. S5.

$$V_{oc-RENPs} \; = \; 91952.32 \; nm^3$$

The RENPs have a hexagonal lattice structure, so the volume of hexagonal unit cell can be calculated as:

$$uV_{hexagonal} = \frac{3\sqrt{3}}{2}a^2c = 0.32\ nm^2$$

where a and c represent the lattice parameters of hexagonal unit cells. The values of the lattice parameters are selected from a previously published article listing a = 5.91 Å and c = 3.53 Å. Therefore, the number of unit cells in an oc-RENP can be calculated as:

$$uN_{hexagonal} = \frac{V_{oc-RENPs}}{uV_{hexagonal}} = 287351$$

During the synthesis of oc-RENPs, the erbium ion doping is set at a concentration of 2%, so the number of Er ions within a single NP is found to be 5747.02. The analysis of the emitting sites from the printed slab utilizing f-RENPs indicates a count of 8.06E16, which contains 4.63E20 Er ions.

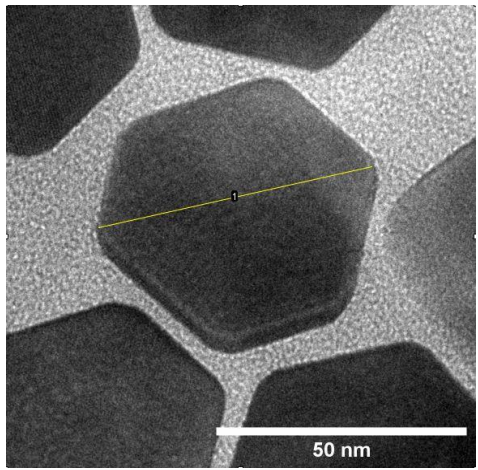

***Fig. S5.*** *Size of single RENPs*

## 6. Dimensions of 2PP printed structures:

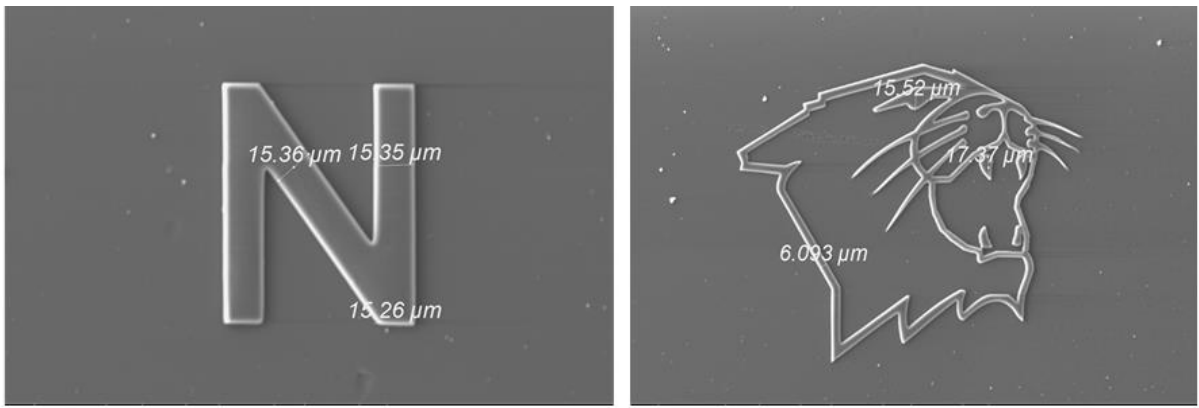

***Figure S6*** *SEM measurements of printed N and wildcat structures. All structures had a height of 1.5 microns.*

**7. Fluorescence images of printed structures using oc-NPs vs. f-NPs.**

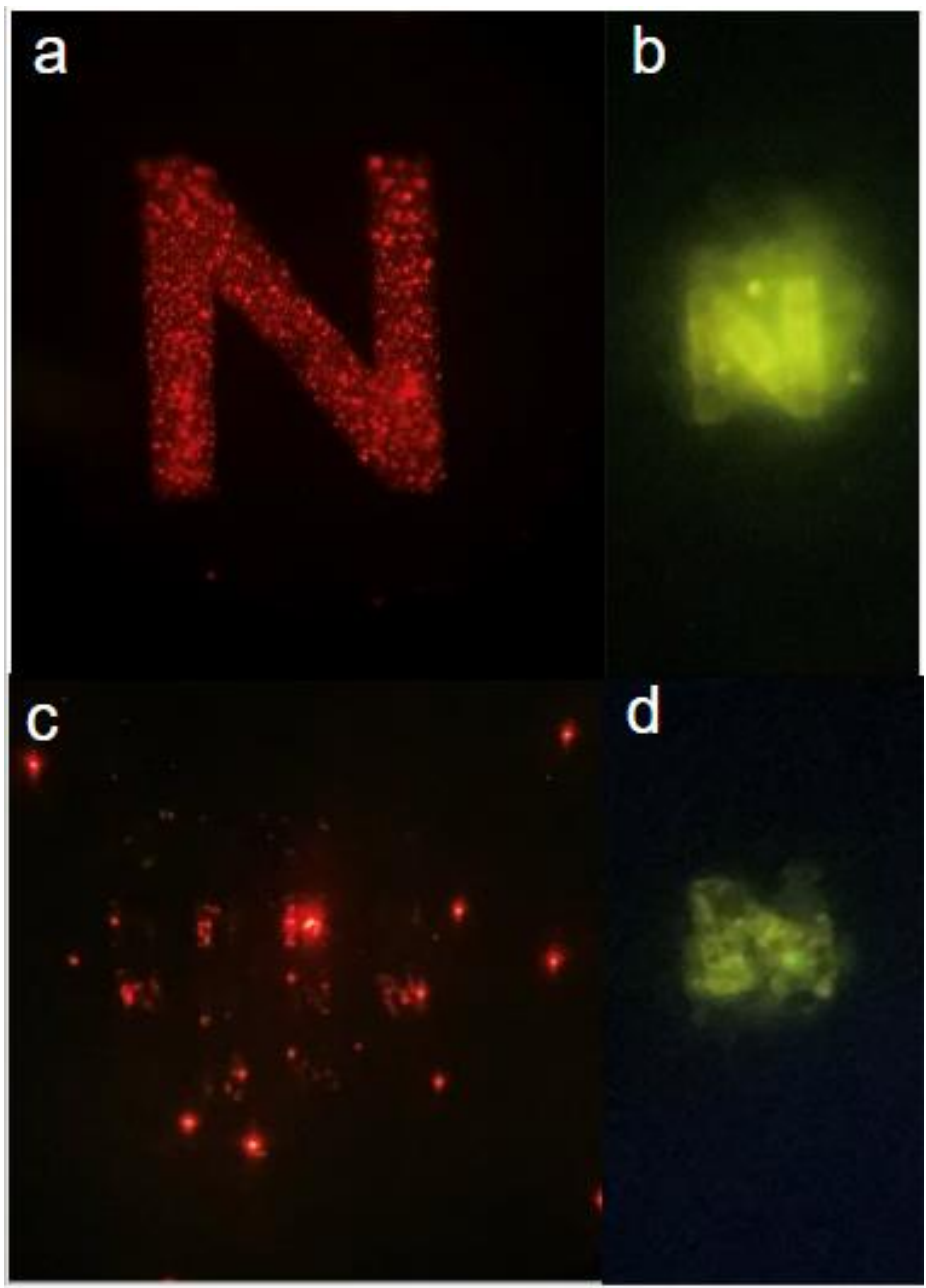

**Figure S7** Fluorescence images of the N structures using QDs and RENPs before and after modification. The f-QDs (a) and the f-RENPs (b) show substantially better loading and dispersion than the oc-QDs (c) and the oc-RENPs (d). In the case of the oc-QDs, multiple N structures were printed, and none achieved a QD loading similar to the f-QD structure.